\documentclass[conference]{IEEEtran}
\IEEEoverridecommandlockouts
\usepackage{cite}
\usepackage{amsmath,amssymb,amsfonts}
\usepackage[utf8]{inputenc}
\usepackage{algorithm} 
\usepackage{algpseudocode}
\usepackage{graphicx}
\usepackage{textcomp}
\usepackage{xcolor}
\usepackage{multirow}
\usepackage{bbm}
\def\BibTeX{{\rm B\kern-.05em{\sc i\kern-.025em b}\kern-.08em
    T\kern-.1667em\lower.7ex\hbox{E}\kern-.125emX}}

\makeatletter
\renewcommand\footnoterule{%
  \kern-3\p@
  \hrule\@width.4\columnwidth
  \kern2.6\p@}
  \makeatother

\usepackage[hang]{footmisc}
\begin{document}

\newcommand{\red}[1]{{\color{red}#1}}
\newcommand{\blue}[1]{{\color{blue}#1}}
\renewcommand{\baselinestretch}{.97}
\renewcommand{\arraystretch}{1.15}

% \title{DynViT: Towards Energy Efficient Dynamic Vision Transformers on Heterogeneous MPSoCs}

%\title{TrimNesS: \underline{\textbf{T}}owards Ene\underline{\textbf{r}}gy-Eff\underline{\textbf{i}}cient Dyna\underline{\textbf{m}}ic \underline{\textbf{N}}eural Networks on H\underline{\textbf{e}}terogeneou\underline{\textbf{s}} MP\underline{\textbf{S}}oCs} % distributed/parallel/mapping 

% \title{Energy Efficient Mapping of Dynamic Neural Networks onto Heterogeneous MPSoCs} 
% \title{Energy Efficient Dynamic Neural Networks Mapping onto Heterogeneous MPSoCs} 
%DynMap
%EMPIRE
%Map and Conquer

\title{Map-and-Conquer: Energy-Efficient Mapping of Dynamic Neural Nets onto Heterogeneous MPSoCs}

% \title{Divide-and-Map: Partitioning Neural Networks for Energy-Efficient Mapping on Heterogeneous MPSoCs}

% \title{Divide-and-Map: Partition Neural Networks for Energy-Efficient Mapping on Heterogeneous MPSoCs}

\author{
\IEEEauthorblockN{
Halima Bouzidi\IEEEauthorrefmark{1}\textsuperscript{\textsection},
Mohanad Odema\IEEEauthorrefmark{2}\textsuperscript{\textsection},
Hamza Ouarnoughi\IEEEauthorrefmark{1},
Smail Niar\IEEEauthorrefmark{1},
Mohammad Abdullah Al Faruque\IEEEauthorrefmark{2}}

\IEEEauthorrefmark{1}\textit{LAMIH/UMR CNRS, Université Polytechnique Hauts-de-France, Valenciennes, France} \\
\IEEEauthorrefmark{2}\textit{Department of Electrical Engineering and Computer Science, University of California, Irvine, USA} \\

\IEEEauthorrefmark{1}\textit{\{firstname.lastname\}}@uphf.fr
\hspace{20truemm}
\IEEEauthorrefmark{2}\textit{\{modema, alfaruqu\}}@uci.edu
\vspace{-5truemm}
}

\maketitle

\begingroup\renewcommand\thefootnote{\textsection}
\footnotetext{Denotes Equal Contribution \\ This work was partially supported by the NSF under award CCF-2140154.}
\endgroup

\begin{abstract}

Heterogeneous MPSoCs comprise diverse processing units of varying compute capabilities. To date, the mapping strategies of neural networks (NNs) onto such systems are yet to exploit the full potential of processing parallelism, made possible through both the intrinsic NNs' structure and underlying hardware composition. In this paper, we propose a novel framework to effectively map NNs onto heterogeneous MPSoCs in a manner that enables them to leverage the underlying processing concurrency. Specifically, our approach identifies an optimal partitioning scheme of the NN along its `width' dimension, which facilitates deployment of concurrent NN blocks onto different hardware computing units. Additionally, our approach contributes a novel scheme to deploy partitioned NNs onto the MPSoC as dynamic multi-exit networks for additional performance gains. Our experiments on a standard MPSoC platform have yielded dynamic mapping configurations that are 2.1x more energy-efficient than the GPU-only mapping while incurring 1.7x less latency than DLA-only mapping.
\end{abstract}

%%Halima: @Mohanad I think we should highlight the novelty in the abstract to the fact that we're incorporating the computation mapping into the design of dynamic neural networks. 


\section{Introduction}
%1) MPSoc what they are, how it is done it today, benefits (brief), composition.
%2) Different componenets offers different performance characterization for different and similar workloads
%3) The problem with conventional neural network mapping is that they underutilize the componenets exisitng in SoCs, and in most cases are deployed on one CUs (GPU/DLA/etc), which is not efficient 
The hardware era has witnessed the emergence of various computing devices, from powerful GPUs to tiny Micro-controllers. To meet the requirements of compute-intensive applications, such as Deep Learning workloads, MPSoCs are designed to incorporate heterogeneous computing units (CU) within the same die, typically sharing the same system memory (DRAM). This hardware architecture paradigm enables the collaborative usage of multiple CUs to accelerate different operations of the same application, hence providing energy savings and performance benefits.
However, the causality between the hardware heterogeneity of MPSoC and the obtained performance for similar and different operations remains an open research question. Indeed, some CUs (e.g., GPUs) can offer high execution speedup at the cost of being energy-hungry, while others, such as NPUs, are power-friendly at the cost of being slow.
Conventional deployment schemes lack a holistic overview of how heterogeneous CUs may behave regarding various computing workloads. In addition, the systematic approach of considering a single CU to deploy an entire application is suboptimal since it overlooks opportunities for further performance gains through maximizing the utilization of the MPSoC's hardware resources. 

Latest research has shed light on the \textit{computation mapping} problem for MPSoC by providing comprehensive modeling methodologies in \cite{song2018sara, monil2020mephesto, xu2021pccs, dagli2022axonn} to characterize computing workloads performances. The resulting models are typically used to map computations onto CUs in a sequential pipeline fashion. However, for workloads exhibiting a high degree of parallelism, such as Neural Networks ($\mathcal{NN}$), there's still room for improvement by refashioning the execution pipeline into parallel stages running concurrently on different CUs, especially considering the inherent capacity for concurrency within $\mathcal{NN}$ layers such as convolutional and multi-head self-attention layers \cite{hadidi2020toward}. 
Prior works \cite{mao2017modnn, shamsa2019goal, hadidi2020toward, hou2022distredge} have considered the computation parallelism on model, data, and task levels. Nevertheless, most works focus on model training rather than inference. Although substantial studies exist for distributed edge devices, few studies have contemplated the case of MPSoCs. 

% Model parallelism serves as an adequate strategy to accelerate single-batch inference. Typically, a $\mathcal{NN}$ is said to be parallelized if its computations can be performed concurrently at certain points. Such a feature is characteristic of numerous common $\mathcal{NN}$ computing layers, such as convolutional, fully-connected, and multi-head self-attention layers \cite{hadidi2020toward}.
%\textcolor{red}{We can delet this section}

%\red{this not further exploit parallelization as this just exploiting dynamism, no? you say in the end that the mapping for parallelism is the unexplored territory}\blue{ HB: Yes, it's an unexplored territory. In S2DNAS the stages are invoked in a sequential manner, thus there's no real parallelism or concurrency}
On the other hand, recent works have started to explore the prospect of partitioning the $\mathcal{NN}$ model itself into separate computing stages that can be invoked in a \textit{dynamic} manner, where simpler inputs can be classified at earlier model stages (i.e., early-exiting), whereas the latter stages are instantiated for more complex inputs.
For instance, S2DNAS \cite{yuan2020s2dnas} demonstrated the benefits from partitioning a model along its width dimension (i.e., layer's channels), and deploying the model as a multi-exit neural network with support for parallelism. Still, studying mapping such parallel neural network components onto a heterogeneous MPSoC for dynamic inference is lacking.

%% HB: We should definitely reduce the text here

% \red{OD: I need to ask an important question about FMAP reuse. we are not 're'-using, these are the actual sizes generated, correct? - How is the memory `reuse' of the GPA and DLA that high???} 
% \blue{HB:They're not actual sizes, just the percentage of concatenations. For GPU and DLA cuz all feature maps are concatenated, so the percentage is 100\%}\red{OD: then 1) this is not reuse 2) this misses the point we wanted to originally emphasize about workloads management in a shared working memory (i.e., how many features do you need to keep at the same time). We'll need to to synchronize how the memory rheotric is addressed throughout the paper}
% \blue{HB: Okay, I understand your point here. Will synchronize the numbers with actual sizes instead of percentage of concatenations. But in this case, should it also computed in a dynamic way? such as in latency and energy? also for baselines, as feature maps are considered as they're originally, they will be entirely written in the shared memory (DRAM), so they're 100\% being used, right?}

% https://uci.zoom.us/j/98516840641?pwd=QmdONTVMeUtMZitrUWUva1lvcW1TQT09

\begin{figure}[ht]
\centering
  \includegraphics[width=0.49\textwidth]{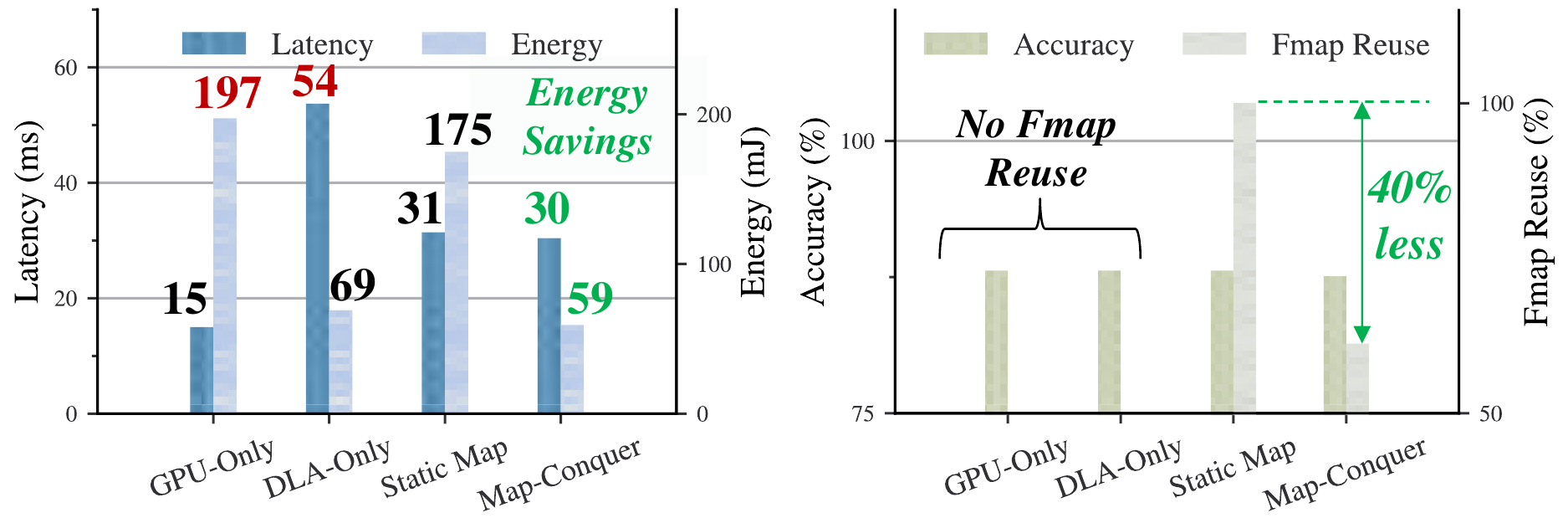}
  \vspace{-4.5ex}
  \caption{Performance comparison between different mapping and deployment options for \textit{Visformer} \cite{chen2021visformer} on Cifar100 and AGX Xavier MPSoC }
  \label{fig:motivation}
  \vspace{-2ex}
\end{figure}

\subsection{Motivational example}

Figure \ref{fig:motivation} illustrates the underlying performance tradeoffs obtained from deploying an $\mathcal{NN}$ onto a heterogeneous MPSoC. Specifically, the example compares different mapping approaches for a \textit{Visformer} architecture \cite{chen2021visformer} 
(from the Vision-Transformers class of $\mathcal{NN}$) 
onto an AGX Xavier MPSoC with a single GPU and two deep learning accelerators (DLAs).  As shown in the left subfigure, mapping the \textit{Visformer} entirely to either hardware computing unit, namely \textit{GPU-Only} and \textit{DLA-Only}, yields a sub-optimal performance: with regards to energy consumption for the former, and with regards to execution latency for the latter. As an alternative, we implemented a \textit{distributed static mapping} strategy that aims to harvest the best of both worlds -- GPU's speed and DLA's energy efficiency. More so, we implement the mapping strategy to exploit the underlying parallelism through \textit{partitioning} \textit{the} \textit{Visformer} along its width dimension (i.e., the attention layer heads), and distributing them along the CUs. Mildly, the \textit{static mapping} strategy leads to performance improvements over its single-mapping counterpart with regards to each component's deficient metric (42.6\% speedup over \textit{DLA-Only} and 11.1\% energy gains over \textit{GPU-only}). Accordingly, we alter our implementation to attain a \textit{dynamic} version of this mapping, namely \textit{Map-Conquer}, 
% \textcolor{red}{At this stage we didn't define what is meant by "dynamic". The reader must have an intuitive and simple definition}\blue{OD: yes; check the following text} 
where the \textit{Visformer} is deployed as a multi-exit neural network on the MPSoC, leading to substantial performance gains due to the nature of dynamic inference. In fact, this \textit{dynamic} mapping strategy dominates the DLA with respect to both the latency (44.4\% speedup) and energy efficiency (14.5\% gain). Still, one deficit from such \textit{distributed} mapping strategies is the additional inter-CU overheads experienced across the MPSoC. In the right sub-figure, we show that adopting a dynamic strategy can also aid in alleviating such burden compared to the static mapping approach. Particularly, our approach identifies the key feature subset from each stage, and only involves those in any needed inter-CUs exchanges, denoted by \textit{Fmap Reuse}. This scheme leads to 40\% less \textit{Fmap Reuse} compared to static mapping (which exchanges all needed features) at the expense of ~0.5\% accuracy drop.   

\subsection{Novel Contributions}
We provide the following novel contributions in this paper 
\begin{itemize} 
    \item We present \textit{Map-and-Conquer}, an energy-efficient execution scheme for Dynamic $\mathcal{NN}$ on MPSoCs. 
    \item We leverage model-parallelism along the ``\textit{width}'' dimension to partition the $\mathcal{NN}$ to multiple inference stages that can be run dynamically and concurrently on the MPSoC.
    \item We derive a comprehensive system model to characterize the performance of the concurrent inference stages on heterogeneous CUs with support for DVFS features.
    \item We design an optimization framework to provide the best partitioning and mapping strategies for Dynamic $\mathcal{NN}$ on the available CUs of the MPSoC.
    \item On the NVIDIA Jetson AGX Xavier MPSoC and various $\mathcal{NN}$ architectures, our experiments demonstrate that \textit{Map-and-Conquer} can achieve up to $\sim$ \textbf{2.1x} more energy-efficiency than the GPU-only mapping while incurring $\sim$ \textbf{1.7x} less latency than DLA-only mapping, all while preserving the desired level of accuracy.
\end{itemize}

\section{Related works}

\subsection{Dynamic Neural Networks}
Dynamic Neural Networks serve as attractive solutions to scale computation according to the input complexity, providing latency speedup and energy gains. Incorporating dynamicity into NN inference has been widely studied for CNN architectures through early-exiting along the architecture's depth \cite{teerapittayanon2016branchynet} or width \cite{yuan2020s2dnas}. Recently, early-exiting is emerging to Vision Transformers (ViT) as they exhibit many computation redundancies \cite{rao2021dynamicvit, yu2022mia}. For instance, MIA-Former \cite{yu2022mia} dynamically adapts the number of heads in attention layers. This latter approach can also be exploited for model partitioning, as it represents the width in ViT. However, most existing works still need to catch the hardware dimension when designing a \textit{dynamic} ViT, which is a vital factor given their complexity.

\subsection{Computation mapping on MPSoCs}
Recent MPSoCs contain diverse heterogeneous CUs that usually share system memory, making them more flexible for collaborative execution. Recent works have explored this specificity of MPSoC to optimize the execution of $\mathcal{NN}$. AxoNN and MEPHESTO \cite{monil2020mephesto, xu2021pccs, dagli2022axonn} propose modeling strategies to characterize execution latency and energy consumption for computation mapping on the AGX Xavier MPSoC. Jedi \cite{jeong2022tensorrt} provides a framework built upon TensorRT to accelerate $\mathcal{NN}$ via model parallelism to maximize throughput for batched inference. \cite{kang2020scheduling, kao2020gamma} proposes evolutionary-based scheduling for NN layers on heterogeneous MPSoCs with \textit{DVFS} by exploiting both data and model parallelism to optimize the throughput. DistrEdge \cite{hou2022distredge} provides a detailed analysis of different model parallelism schemes for distributed computing over edge devices. However, none of the prior works have considered the design of dynamic NN in the computation mapping problem for collaborative execution on MPSoCs.

To the best of our knowledge, our work is the first to address the problem of dynamic $\mathcal{NN}$ design and mapping onto heterogeneous MPSoC in a collaborative manner. Thus exploiting $\mathcal{NN}$ dynamicity, MPSoC heterogeneity, and reconfigurability (DVFS) for an energy-efficient execution on MPSocS. Table \ref{tab:sota_table} highlights the key differences between related works and Ours.

\vspace{-2ex}

\begin{table}[ht!]
\centering
\caption{Comparison between Related-works and ours 
} 
\fontsize{9}{9}\selectfont
\scalebox{0.75}{
\label{tab:sota_table}
\begin{tabular}{cccccc} 
\hline
Related Work                                           & \begin{tabular}[c]{@{}c@{}}Early\\Exiting\end{tabular} & \begin{tabular}[c]{@{}c@{}}Model \\Parallelism\end{tabular} & \begin{tabular}[c]{@{}c@{}}Collaborative \\execution\end{tabular} & DVFS       & \begin{tabular}[c]{@{}c@{}}Training\\free\end{tabular}  \\ 
\hline
AxoNN \cite{dagli2022axonn}           &                                                        &                                                             & x                                                                 &            & x                                                       \\
Jedi \cite{jeong2022tensorrt}         &                                                        & x                                                           & x                                                                 &            & x                                                       \\
DistrEdge \cite{hou2022distredge}     &                                                        & x                                                           & x                                                                 &            & x                                                       \\
Kang et al. \cite{kang2020scheduling} &                                                        & x                                                           & x                                                                 & x          & x                                                       \\
S2DNAS \cite{yuan2020s2dnas}          & x                                                      & x                                                           &                                                                   & x          &                                                         \\
HADAS \cite{bouzidi2023hadas}     & x                                                      &                                                             &                                                                   & x          &                                                        \\
Edgebert \cite{tambe2021edgebert}     & x                                                      &                                                             &                                                                   & x          & x                                                       \\ 
\hline
\textbf{Ours}                                          & \textbf{x}                                             & \textbf{x}                                                  & \textbf{x}                                                        & \textbf{x} & \textbf{x}                                              \\
\hline
\end{tabular}
}
\vspace{-2ex}
\end{table}

%% HB: I think we should use one term for the MPSoC, in this case, I'm gonna replace heterogeneous SoC --> with MPSoC, this will save space too
\section{System Model}
In this section, we model the components needed to conduct a static-to-dynamic transformation of $\mathcal{NN}$, and characterize its performance overheads when executing on the heterogeneous MPSoC accordingly.

\subsection{Dynamic Transformation of NNs on MPSoC}
\label{subsec:stat_to_dyn}
\begin{figure}
    \centering
    \includegraphics[width=0.4\textwidth]{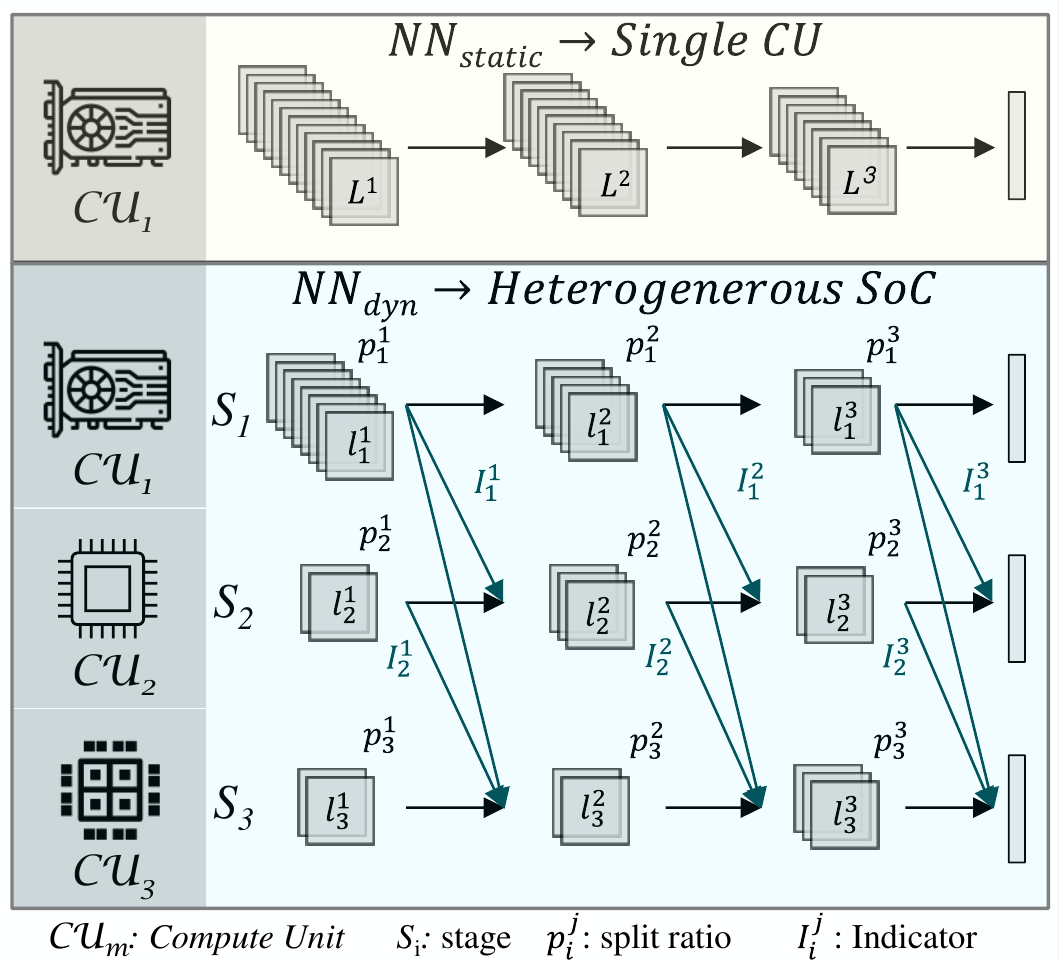}
    \vspace{-1ex}
    \caption{Transformation of $NN_{static}$ into $NN_{dyn}$ based on $s$ and $I$, and mapping $NN_{dyn}$ onto a MPSoC with multiple $\mathcal{CU}$s}
    \label{fig:DyNN}
    \vspace{-2ex}
\end{figure}

Consider an unaltered basic neural network, $\mathcal{NN}$, constituting a sequence of $n$ computational layers as follows:
\begin{equation}
    \mathcal{NN} = \mathcal{L}^{n} \circ \mathcal{L}^{n-1} \circ ... \circ \mathcal{L}^{1} \label{eqn:NN}
\end{equation}
in which each computing layer, $L^j$, consists of weight parameter matrices whose count represents the `width' of the layer. Without losing generality, we refer to these weight matrices here as `channels', such as those in a convolutional $\mathcal{NN}$. Therefore, we can define the $j^{th}$ layer as:
\begin{equation}
    \mathcal{L}^j = \{C_1^j, C_2^j, ..., C_W^j\} \label{eqn:channels}
\end{equation}
in which $C_i^j$ represents the $i^{th}$ channel in the $j^{th}$ layer. Now, consider an SoC that comprises $M$ computing units $\mathbbm{CU} = \{\mathcal{CU}_1, \mathcal{CU}_2, ..., \mathcal{CU}_M\}$, the goal is to devise a strategy to partition every $L^j$ into $M$ subsets according to its width dimension (i.e., the channels), and thus, $\mathcal{L}^j$ is redefined as:
\begin{equation}
    \mathcal{L}^j = \{l_1^j, l_2^j, ..., l_M^j\} \label{eqn:l_small}
\end{equation}
which enables every contiguous subset of channels, $l_m^j$, to be mapped onto one of the computing units, $\mathcal{CU}_m \in \mathbbm{CU}$. In this sense, we define two operations to characterize this mapping problem: (\emph{i}) \textit{\textbf{Partitioning}}; to divide layers and generate the subsets $l_m^j$, and (\emph{ii}) \textit{\textbf{Concatenation}}; to reuse the generated intermediate features, $F_m^{j}$, in set of the immediate next layer in all subsequent stages, \{$l_{m+1:M}^{j+1}$\}. In accordance, we define two parameter matrices to characterize these operations:
\begin{equation}
    \mathbbm{P} = \begin{bmatrix}
            p_1^1 & \cdots & p_1^n\\
            \vdots & \ddots & \vdots\\
            p_M^1 & \cdots & p_M^n
        \end{bmatrix} ,\; \mathbbm{I} = \begin{bmatrix}
            I_1^1 & \cdots & I_1^n\\
            \vdots & \ddots & \vdots\\
            I_M^1 & \cdots & I_M^n
        \end{bmatrix} \label{eqn:mtx} \;
\end{equation}
where $\mathbbm{P}$ is the \textit{partitioning matrix} in which every $p_i^j$ represents the fraction of channels in a layer $L^j$ (equation \ref{eqn:channels}) are to be assigned to $l_i^j$. $\mathbbm{I}$ is an \textit{indicator matrix} in which $I_i^j \in \{0,1\}$ indicates whether the intermediate features, $F_i^{j}$, are to be used in the $j+1$ layers in the following stages. Figure \ref{fig:DyNN} provides an illustration for how these matrices govern the partitioning and concatenation operations of a neural network. As shown, each $\mathcal{CU}_m$ on the SoC can host a unique sequence of channel subsets, which we denote as a stage, $S_i$, given as:
\begin{equation}
    S_i = l_i^n \circ l_i^{n-1} \circ ... \circ l_i^{1}
\end{equation}
and ultimately, we obtain the following set of stages:
\begin{equation}
    \mathbb{S} = \{S_1, S_2, ..., S_M\}
\end{equation}
if we augment each stage $S_i$ with an exit at its tail (e.g., a classifier layer), each stage can now act as a \textit{separate} inference sub-model, to be invoked based on some established runtime criteria during deployment (e.g., input processing difficulty).

Lastly, we define an additional vector, $\mathbb{M}$, to parameterize the mapping of stages onto the SoC: $S_i\rightarrow \mathcal{CU}_m\ \forall\ S_i \in \mathbb{S}, \mathcal{CU}_m \in \mathbbm{CU}$. $\mathbb{M}$ can by given as:
\begin{equation}
    \mathbb{M} = [\pi_1, \hdots, \pi_M]\;\; s.t.\; \pi_k \neq \pi_{k'} \;\forall\; 1\leq k \leq k' \leq M
\end{equation}
in which every entry $\pi_k$ is one $\mathcal{CU}_m \in \mathbbm{CU}$ to whom $S_k$ is mapped. The condition is for enforcing that no two stages are mapped onto the same $\mathcal{CU}_m$. 

% With that, we have modeled a `dynamic' inference model on heterogenous SoCs and in the following subsection, we characterize its processing overheads on the different hardware components and show how we can leverage their separation to attain a parallel execution model.

\subsection{Distributed Performance Modelling for Dynamic Inference}
\label{subsec:sys_model}
Here, we model the dynamic inference execution overheads given the partitioned deployment of a model on a heterogeneous MPSoC with regards to \textit{latency} and \textit{energy consumption}. 
Given the scope of this work, we assume \textit{ideal input mapping} in which the number of stages needed to process an input sample $i$ is known apriori. In practice, input mappings can be determined using runtime controllers as those stated in \cite{bouzidi2023hadas}.

\smallskip
\noindent
\textbf{Execution Latency.} Let $\tau_i^j$ denotes the execution latency overhead of sublayer $l_i^j$ in $S_i$. We first aim to derive an expression for the latency overhead of every stage, denoted by $T_{S_i}$. At this point, we highlight that stages are indexed by the order of their execution. For example, $S_2$ is only instantiated if $S_1$ is deemed insufficient to terminate the processing. Thus, there exists inter-stage dependencies of $S_i$ on its predecessors $S_{1:i-1}$ (as indicated by $I_i$) whose overheads need to be accounted for, especially when stages are mapped onto different hardware units. 
\begin{figure}
    \centering
    \includegraphics[width=0.48\textwidth]{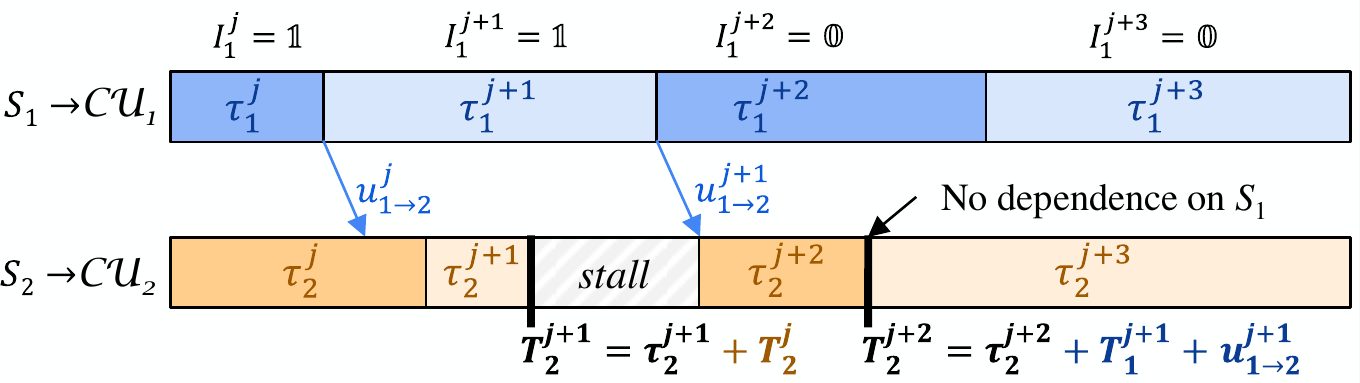}
    \vspace{-1.5ex}
    \caption{Concurrent execution of $S_2$ and $S_1$ considering timing dependencies}
    \label{fig:SoC_timeline}
    \vspace{-2ex}
\end{figure}
% Trivially, such dependencies can be managed through a sequential execution model of stages, where intermediate results are passed to subsequent stages if needed. This scheme, however, can \textit{undo} any potential speedups gained from adopting dynamic inference in the first place, especially when more input samples are assigned to the latter stages. 
To avoid the demerits of a sequential execution model, we leverage the underlying separation of the compute units and propose a \textit{concurrent} model of execution that considers these dependencies. Specifically, any sublayer $l_i^j$ in an `instantiated' $S_i$ can immediately proceed to execute its inputs once all of its required input features, \{$(F_{1:i-1}^{j-1}\cdot I_{1:i-1}^{j-1}) \cup F_i^{j-1}$\}, are readily available within its local vicinity. From here, we can give the \textit{cumulative} latency overhead at $l_i^j$ by:
\begin{equation}
    T_i^j = \tau_i^j + \max\{T_i^{j-1}, T_k^{j-1} + u_{k\rightarrow i}^{j-1}\;|\;I_{k}=\mathbbm{1}\; \forall\; 1 \leq k < i\}
    % (T_i^{j-1}, \sum_{k=1}^{i-1} T_k^{i-1}
\end{equation}
where the second term captures the maximum cumulative latency experienced in a previous layer from all stages preceding $S_i$. Thus, $T_i^{j}$ captures the cumulative latency estimate in stage $i$ at $j$ while accounting for inter-stage dependencies, while $u_{k\rightarrow i}^{j-1}$ is the data transmission overhead of the features $F_{k}^{j-1}$ to the local buffer of the computing resource assigned to $S_i$ (See Figure \ref{fig:SoC_timeline} for an illustrative example). Given $n$ layers in $S_i$, the execution latency of $S_i$ can be estimated:
\begin{equation}
    T_{S_i} = T_i^n \label{eqn:latency_stage}
\end{equation}

\smallskip
\noindent
\textbf{Energy Consumption.} For every $\mathcal{CU}_m \in \mathbbm{CU}$, we first characterize its power consumption as follows:
\begin{equation}
    P_m = P_m^s + P_m^d(\vartheta_m) \approx \alpha + \beta\cdot\vartheta_m
\end{equation}
$P_m^s$ and $P_m^d$ are the static and dynamic components, respectively. The latter is parameterized by the scaling factor $\vartheta_m$ based on the supported DVFS features on $\mathcal{CU}_m$, where $\alpha_m$ and $\beta_m$ are constants. From here, the energy required to complete processing at sublayer $l_i^j$ during inference is given by:
\begin{equation}
    e_i^j = \tau_i^j \cdot P_m
\end{equation}
and as such the total energy consumed by $S_i$ is:
\begin{equation}
    E_{S_i} = \sum_{j=1}^n e_i^j \label{eqn:energy_stage}
\end{equation}

\smallskip
\noindent
\textbf{Overall Characterization.} Under the concurrent model of execution, the overall performance characterization is given by the following two equations: 
\begin{equation}
    T_{\mathbbm{P}, \mathbbm{I}, \mathbb{M}, \vartheta} = \max\{T_{S_i}\; \forall \; S_i\in \mathbb{S}\}
    \label{eqn:latency}
\end{equation}
\begin{equation}
    E_{\mathbbm{P}, \mathbbm{I}, \mathbb{M}, \vartheta} = \sum_{i=1}^{M'} E_{S_i}\; s.t. \; 1 \leq i \leq M' \leq M   
    \label{eqn:energy}
\end{equation}
where for a dynamic inference on a MPSoC, described through the parameters choices of ($\mathbbm{P}, \mathbbm{I}, \mathbb{M}, \vartheta$), its execution latency is the \textit{maximum} from all its stages due to concurrency, whereas its energy consumption is the \textit{aggregation} of energy consumed by the $M'$ `instantiated' stages to process an input sample.
% The decision of input instantitation is a topic on its own and is out of the scope of this paper.

\begin{figure}
    \centering
    \includegraphics[width=0.44\textwidth]{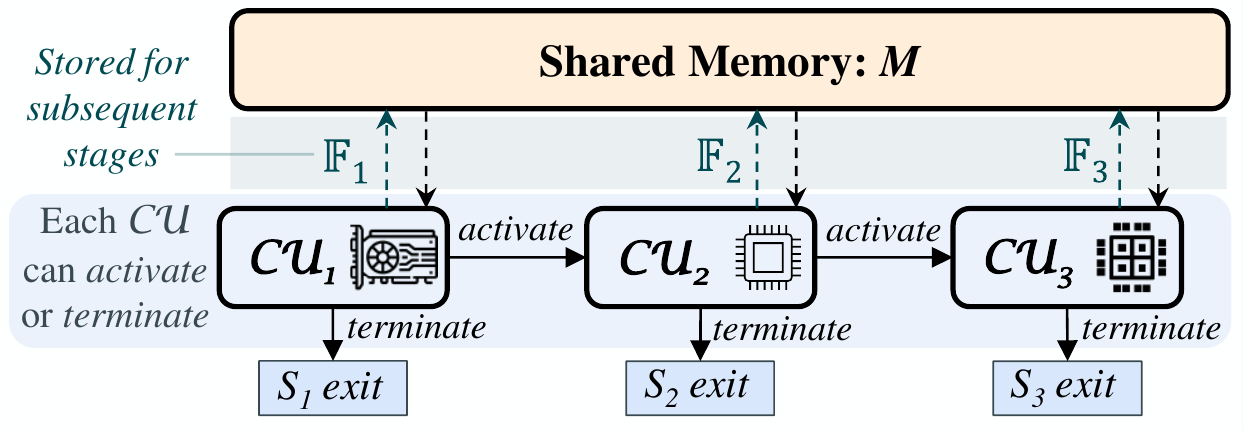}
    \vspace{-2.5ex}
    \caption{Illustration of data movement and feature storage on the MPSoC}
    \label{fig:memory}
    \vspace{-3ex}
\end{figure}

\section{Problem Formulation}

% Let $\Pi(\mathbbm{P}, \mathbbm{I}, \mathbb{M}, \vartheta)$ constitute the mapping of every stage $S_i \in \mathbb{S}$ to a computing unit $\mathcal{CU}_m\in \mathbbm{CU}$, parameterized by the choices of $\mathcal{NN}$'s splitting and concatenation, as well as the current DVFS settings on each viable hardware. Our goal is to find an ideal mapping policy that can maximize performance efficiency under a set of constraints as follows:

Let $\Pi = (\mathbbm{P}, \mathbbm{I}, \mathbb{M}, \vartheta)$ combine all parameters that characterize a neural network's mapping onto an MPSoC. Our main optimization goal is to find the ideal parameters that can enhance a performance objective, $\mathcal{P}$, given a set of constraints:
\begin{equation}
   \Pi^* = \min_{\Pi} \mathcal{P}(\Pi) \label{eqn:obj}
\end{equation}
\begin{equation*}
    s.t.\  T_{\Pi^*} < T^{TRG}, \;\; E_{\Pi^*} < E^{TRG}, \;\; \mathbbm{size}_{\Pi^*}(\mathbb{F, I}) < M \;\; \label{eqn:constraints}
\end{equation*}
where $ T^{TRG}$ and $E^{TRG}$ are the respective target latency and energy constraints as set by the practitioner. The constraint $\mathbbm{size}_{\Pi}(\mathbb{F, I}) < M$ is to bound the size of the intermediate features that need to be made readily available for the duration of the inference (denoted as $\mathbb{F}$), for they are limited by the MPSoC's shared memory size, $M$ (see Figure \ref{fig:memory}). $\mathcal{P}$ is kept generic and can be tuned to the designers' objectives.

\section{Proposed Framework}
\begin{figure}
    \centering
    \includegraphics[width=0.48\textwidth]{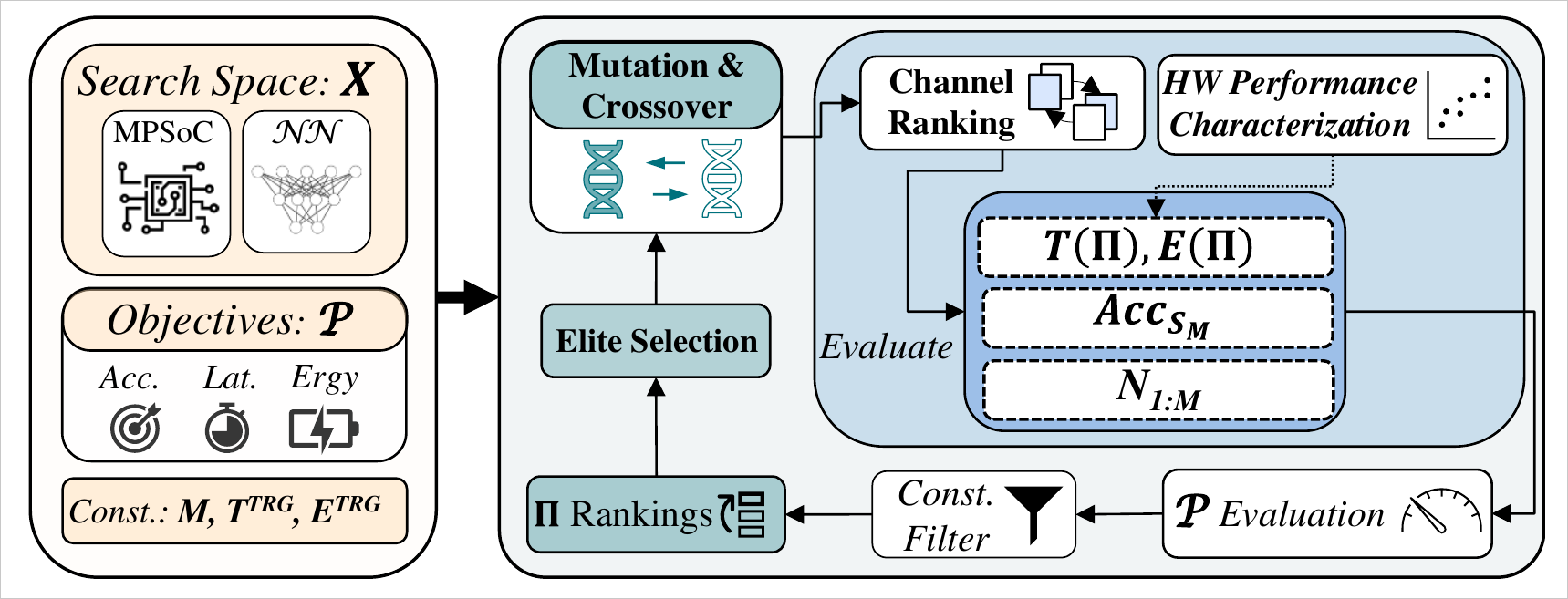}
    \vspace{-1.5ex}
    \caption{Overview of our proposed optimization framework }
    \label{fig:overview}
    \vspace{-3ex}
\end{figure}

In this section, we propose an optimization framework to solve the mapping problem. Figure \ref{fig:overview} gives an overview of our framework, whose key components are detailed below.

\subsection{Search Space} 
% Need to revise the complexity of the search space
%% HB: I think we've already named the search space in the problem formulation, instead of X, we're gonna just use $\Pi$

Here we describe how to generate a search space, $X$ of mapping strategy parameters, namely the space of  ($\mathbb{P}, \mathbb{I}, \mathbb{M}, \vartheta$). Firstly,  given a pretrained $\mathcal{NN}$ and an MPSoC with $M$ CUs, we can generate $X$ based on the $\mathcal{NN}$'s layer specifications and the MPSoC's underlying hardware composition. For the former, the attainable depth and width parameters of every layer $L^j \in \mathcal{NN}$ define the $(\mathbbm{P}, \mathbbm{I})$ parameter matrices. For the latter, $M=|\mathbbm{CU}|$ specifies its mapping space and the total number of inference stages. Lastly, $\vartheta$ is specified through the hardware reconfiguration parameters (\textit{DVFS}). For instance, the mapping search space complexity of one layer from the \textit{Visformer} \cite{chen2021visformer} is $\mathcal{O}(1.5 \times 10^5)=\mathcal{O}(8^3 \times 3! \times 50)$, considering $8$ channel partitioning ratios, $M=3$, and $|\vartheta|=50$. %\red{$w$ not defined}

%To analyze the search space complexity, we initially consider the mapping of one layer $\mathcal{L}^{j}$ with $w$ possible partitioning of channels across $M$ CUs, which yields a complexity of becomes $\mathcal{O}(w^M \times 2^M)$. Now, If we consider $w$ is the same across every layer, the overall complexity for $N$ model layers becomes $\mathcal{O}((w^M \times 2^M)^N \times M! \times |\vartheta|$). Here, $N$ exponentially impacts the complexity of the search space and makes it challenging to explore. Note that for any Neural Network, $N$ defines the number of \textit{possible-to-parallelize} layers.

\subsection{Performance Objectives}
Next, a performance objective needs to be designated as $\mathcal{P}$ for the main optimization function in equation (\ref{eqn:obj}), to be specifically used for the candidate mapping evaluation. For our case, we used the following expression for $\mathcal{P}$: 
\begin{equation}
    \mathcal{P} = (\frac{Acc_{base}}{Acc_{S_{M}}}) \times (\sum_{i=1}^M T_{S_i}\cdot N_i) \times (\sum_{i=1}^M E_{S_{1:i}}\cdot N_i) \label{eqn:obj_search}
\end{equation}
In which $Acc_{base}$ is the baseline accuracy of the pretrained $\mathcal{NN}$ model; $Acc_{S_{M}}$ is the accuracy of the last stage of the dynamic version of $\mathcal{NN}$ as its base accuracy. The aforementioned terms are included to ensure that no accuracy drops ensue when a model's structure changes through the $\mathbb{I}$ matrix. $N_i$ represents the number of input samples -from the validation dataset- correctly classified at $S_i$, given that every prior stage misclassifies them. $T_{S_i}$ is the latency experienced by the MPSoC at stage $S_i$ based on equation (\ref{eqn:latency_stage}); $E_{S_{1:i}}$ is the energy consumed by the system as the result of executing $i$ stages of the model -- each $E_i$ is evaluated as in equation (\ref{eqn:energy_stage}).
% This formulation enables us to characterize the performance of the dynamic version of the model compared to its static baseline. 

% Constraint are provided by the user

% Here we aim to maintain the accuracy of the baseline $\mathcal{NN}$ $Acc_{base}$ in the last inference stage by maximizing $Acc_{S_{M}}$. Latency and energy gains are weighted by the ratio of correctly classified samples $N_i$ at each inference stage. Finally, $\omega_1$, $\omega_2$, and $\omega_3$ are user-defined relative importance weights.

\subsection{Search Algorithm} 
%the steps of the optimization algorithm: from dynamic neural network design to hardware mapping, with channels ranking...etc. The encoding and sampling mechanism? How we apply mutation and crossover?
% No need to mention additional details on the sampling, encoding,..etc. As we discussed earlier, we present the search algo as a tool
%Will talk abou channels ranking in the next sub-section as it's part of the evaluation and not the search algo itself

% Regarding the complexity of the search space, conventional methods (e.g., exact methods, random search) are unable to provide optimal solutions even with high optimization budgets. 
%% HB: I think no need to mention NSGA-II, we can keep it a general evolutionary algorithm
We develop an evolutionary-based algorithm to effectively explore the search space. Following the workflow in Figure $\ref{fig:overview}$, every new search iteration entails a new population, say $X'_{i}\subset X$. Then for every configuration $\Pi \in X'$, its corresponding dynamic $\mathcal{NN}$ and hardware settings are evaluated using a predefined objective function, $\mathcal{P}$. Based on results, configurations that do not meet the search constraints (e.g., memory usage) are omitted, whereas the remaining ones are ranked according to $\mathcal{P}$, and a subset of elite configurations is taken for a mutation and crossover stage to obtain the new population $X'_{i+1}$. Once the search budget expires, a Pareto set in calculated from all the generated populations from which the ideal dynamic mapping strategy is extracted. 
%the corresponding population is taken as the final solution from which the ideal dynamic mapping strategy is extracted. --> We extract the Pareto set from all the generated populations and not only the last one

% Figure \ref{fig:overview} details the algorithm’s workflow. First, we sample a random population from $\Pi$ then we select the elite individuals based on the feedback from the objective evaluation of $\mathcal{P}$. Afterward, we create the child population by applying mutation and crossover on the decision variables of $\mathbbm{P}, \mathbbm{I}, \mathbb{M}$, and  $\vartheta$, separately but with similar probabilities to ensure a fair and balanced exploration between the different components. Finally, this optimization cycle is repeated until reaching the maximum budget defined in the search specifications \textit{(aka. Search param.)}.

\subsection{Channel Partitioning and Reordering}
Before a candidate configuration $\Pi\in X'$ is evaluated on the objective function $P$, the $\mathcal{NN}$ should be partitioned according to the ratios in $\mathbb{P}$. Yet to maximize performance when partitioning, the width channels in each model layer are arranged according to their \textit{degree of importance}. The logic being that given the sampled partitioning matrix $\mathbb{P}$ for a configuration $\Pi$, it would be beneficial to assign the most important channels in the layer to the earlier inference stages for dynamic inference. This would enable numerous samples to terminate processing prematurely if deemed feasible, which will consequently aid in enhancing the \textit{dynamic inference} performance of the $\mathcal{NN}$ with regards to experienced latency and energy on the MPSoC. This reordering method is feasible as all channels within the same layer share the same dimensions. Channel ranking is widely used for network pruning, and we follow the approach in \cite{molchanov2019importance} to estimate each channel's importance.

\begin{table}[t!]
\centering
\caption{Performances Breakdown of the Pareto optimal models obtained by Map-and-Conquer and the baselines} 
% \fontsize{9}{9}\selectfont
\scalebox{0.75}
{
\label{tab:perf_breakdown}
\begin{tabular}{cccccc} 
\hline
\begin{tabular}[c]{@{}c@{}}\textbf{Opt.}\\\textbf{Strategy}\end{tabular}      & \begin{tabular}[c]{@{}c@{}}\textbf{NN}\\\textbf{Implment.}\end{tabular} & \begin{tabular}[c]{@{}c@{}}\textbf{\textbf{TOP-1 Acc }}\\\textbf{\textbf{(\%)}}\end{tabular} & \begin{tabular}[c]{@{}c@{}}\textbf{\textbf{Avg. Enrg. }}\\\textbf{\textbf{(mJ)}}\end{tabular} & \begin{tabular}[c]{@{}c@{}}\textbf{\textbf{Avg. Lat. }}\\\textbf{\textbf{(ms)}}\end{tabular} & \begin{tabular}[c]{@{}c@{}}\textbf{Fmap. reuse.}\\\textbf{(\%)}\end{tabular}  \\ 
\hline
\multicolumn{6}{c}{\textbf{Visformer (ViT-based Architecture)}}                                                                                                                                                                                                                                                                                                                                                                                                                                                                       \\ 
\hline
\multirow{2}{*}{None}                                                         & GPU                                                           & \multirow{2}{*}{\textbf{88.09}}                                                              & 197.35                                                                                        & \textbf{15.01}                                                                               & -                                                                          \\
                                                                              & DLA                                                           &                                                                                              & \textbf{69.22}                                                                                & 53.71                                                                                        & -                                                                           \\ 
\hline
\multirow{2}{*}{\begin{tabular}[c]{@{}c@{}}No Fmap\\Constr.\end{tabular}}   & Ours-L                                                                  & 86.12                                                                                        & \textbf{108.44}                                                                               & 25.58                                                                                        & 68.75                                                                         \\
                                                                              & Ours-E                                                                  & \textbf{87.58}                                                                               & \textbf{59.21}                                                                                & \textbf{30.40}                                                                               & \textbf{61.25}                                                                \\ 
\hline
\multirow{2}{*}{\begin{tabular}[c]{@{}c@{}}75\% Fmap\\Constr.\end{tabular}} & Ours-L                                                                  & 84.64                                                                                        & 102.67                                                                                        & 24.65                                                                                        & 65.00                                                                         \\
                                                                              & Ours-E                                                                  & \textbf{87.67}                                                                               & \textbf{65.12}                                                                                & \textbf{29.46}                                                                               & 75.00                                                                         \\ 
\hline
\multirow{2}{*}{\begin{tabular}[c]{@{}c@{}}50\% Fmap\\Constr.\end{tabular}} & Ours-L                                                                  & 82.69                                                                                        & 116.00                                                                                        & \textbf{24.51}                                                                               & 50.00                                                                         \\
                                                                              & Ours-E                                                                  & 84.16                                                                                        & 82.44                                                                                         & 32.70                                                                                        & \textbf{50.00}                                                                \\ 
\hline
\multicolumn{6}{c}{\textbf{VGG19 (CNN-based Architecture)}}                                                                                                                                                                                                                                                                                                                                                                                                                                                                           \\ 
\hline
\multirow{2}{*}{None}                                                         & GPU                                                               & \multirow{2}{*}{\textbf{80.55}}                                                              & 630.11                                                                                        & \textbf{25.23}                                                                               & -                                                                           \\
                                                                              & DLA                                                               &                                                                                              & \textbf{164.89}                                                                               & 114.41                                                                                       & -                                                                           \\ 
\hline
\multirow{2}{*}{\begin{tabular}[c]{@{}c@{}}No Fmap\\Constr.\end{tabular}}   & Ours-L                                                                  & 84.81                                                                                        & 251.63                                                                                        & \textbf{25.67}                                                                               & 52.94                                                                         \\
                                                                              & Ours-E                                                                  & 84.63                                                                                        & 153.97                                                                                        & \textbf{34.02}                                                                               & 70.58                                                                         \\ 
\hline
\multirow{2}{*}{\begin{tabular}[c]{@{}c@{}}75\% Fmap\\Constr.\end{tabular}} & Ours-L                                                                  & 84.76                                                                                        & \textbf{247.34}                                                                               & \textbf{26.07}                                                                               & 64.70                                                                         \\
                                                                              & Ours-E                                                                  & 82.64                                                                                        & \textbf{136.31}                                                                               & 37.22                                                                                        & \textbf{47.05}                                                                \\ 
\hline
\multirow{2}{*}{\begin{tabular}[c]{@{}c@{}}50\% Fmap\\Constr.\end{tabular}} & Ours-L                                                                  & 84.62                                                                                        & 250.80                                                                                        & 25.83                                                                                        & 50.00                                                                         \\
                                                                              & Ours-E                                                                  & 82.53                                                                                        & \textbf{136.41}                                                                               & 37.24                                                                                        & \textbf{50.00}                                                                \\
\hline
\end{tabular}
}
\vspace{-3ex}
\end{table}

\subsection{Performance Evaluation}
%\begin{itemize}
%    \item channels ranking before the split --> should mention that
%    \item Layerwise performance modeling for latency and energy
%    \item the prediction features are composed of: layer configuration, type of the hardware unit, DVFS policy.
%    \item I should add the details on the performance characterization step: how latency and energy were measured? especially for DLA, need to mention that input/output data transmission from system memory to NVM is included in the latency+energy measurements. 
%\end{itemize}

Once a model is transformed to its dynamic version through $\mathbb{P}$ and $\mathbb{I}$, the hardware measurements needed for the performance evaluation of each $\mathcal{NN}$ in equation (\ref{eqn:obj_search}) need to be estimated for each input sample. One way to achieve this is through surrogate models, which are able to predict $\tau_i^j$ and $e_i^j$ of each layer $j$ mapped onto stage $i$ (also CU $i$) based on input configurations while abiding by any inter-stage execution dependencies, and taking into account the computation cost and feature map communication overheads.
Hence, a predictor (XGBoost \cite{chen2016xgboost} in our case) 
is first trained on a benchmarked dataset of diverse layer specifications, deployment hardware and DVFS settings. Afterwards, the predictor is deployed to characterize the performance of each model sampled within the population, providing estimates for its base latency, $\tau_i^j$, and energy consumption, $e_i^j$. In our case, we use the TensorRT library to first evaluate performance overheads on a layer-wise granularity, construct the dataset, and then deploy the predictor to provide hardware evaluations to involved models.

\section{Experiments}

\subsection{Experimental Setup}
Our experiments are conducted on the MPSoC provided by NVIDIA: \textit{Jetson AGX Xavier}. This platform embeds CPU, GPU, and DLA cores on the same chip, sharing the same system memory. To run the $\mathcal{NN}$ workloads on the DLA, we use TensorRT and ONNX to build inference engines from the PyTorch model. As $\mathcal{NN}s$, we use \textit{Visformer} \cite{chen2021visformer} as ViT-based architecture and \textit{VGG19} \cite{simonyan2014very} as CNN-based architecture to validate our approach for both cases. The dataset used for accuracy assessment is CIFAR100. 
Regarding the optimization framework, we run the optimization algorithm for 200 generations, each with a population size of 60, resulting in 12K overall evaluations. Furthermore, the evaluation step is performed on a cluster of 12 GPUs taking up to $\sim$ 10 GPU hours to run the entire optimization process.
%\blue{we assume no cache tx costs $u$, and $I_k$ is the same for all receivers}
%\blue{if space allows, should add something about DVFS}

\subsection{Search Process Analysis} 
%\blue{1) only analyze the search process for the 3 visformer cases. REMOVE VGG PARETO FIGURE}
%\blue{2) Paragraph sizes do not need to be super big (more or less the same as Hadas). Just emphasize on the key findings and differences between the 3 figures - e.g., the annotations, how the accuracy starts to drop in constrained searches, and how energy-oriented models go D-G-G and latency-oriented ones go G-D-D can be mentioned here}
%\blue{3) Writing can be along the lines: "In this part, we analyze the search processes conducted by our framework under 2 cases, one for unconstrained optimization for energy and latency, whereas the other under memory constraints. In Figure \ref{pareto}, we illustrate the explored models alongside under 3 cases: unconstrained, 75\%, and 50\%. As shown, we find ... "}
%\blue{4) \textbf{An idea for motivational figure:} take a1 model, and compare against gpu only, dla only, and its static version with regards to energy, latency, and memory (fmap reuse). (one subfigure for the first two metrics and another for the latter two). In the text, you can say that the max accuracy drop is 0.5\%.  Talk to me}

\begin{figure*}[!ht]
\begin{center}
{\includegraphics[,width = .98\textwidth]{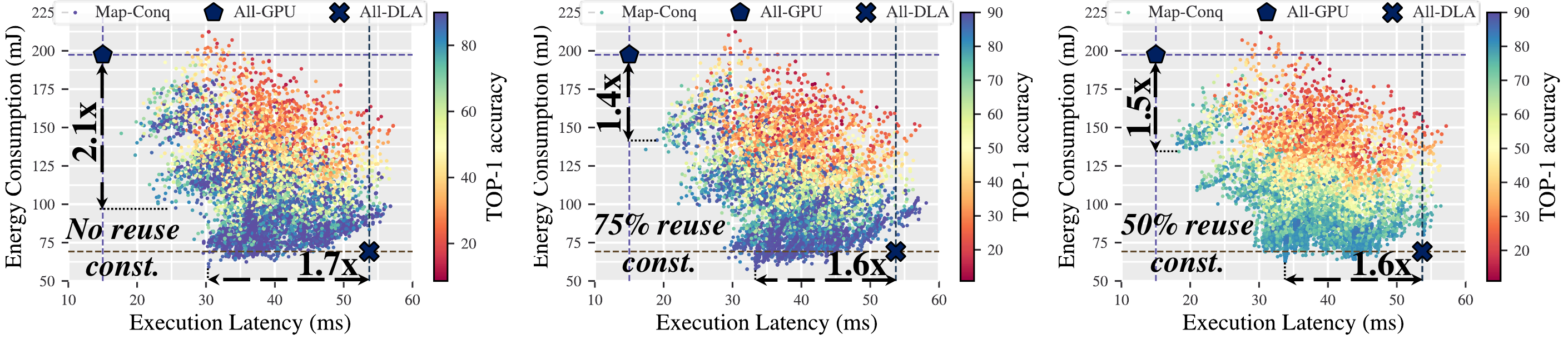}}
\end{center}
\vspace{-3ex}
\caption{Results of three different search strategies: \textbf{Left)} No constraint is set on the \textit{Fmap Reuse}. \textbf{Middle)} Under a constraint of reusing only less than 75\% of feature maps. \textbf{Right)} Under a constraint of reusing only less than 50\% of feature maps. All the results are reported for \textit{Visformer} on the AGX Xavier MPSoC. In the three plots, we highlight the configurations that exhibit the highest latency-energy tradeoff while preserving less than 0.5\% drop in accuracy
} 
\label{fig:opt-results}
\vspace{-3ex}
\end{figure*}

% Explain the Fmaps reuse metric
In this section, we analyze the results of the search process conducted by our framework under two main cases: 1) When no constraint is set to limit the feature map reuse between inference stages, 2) When only less than 75\%, 50\% of feature maps can be reused, respectively. 
In Figure \ref{fig:opt-results}, we show the optimization results for each case. Firstly, we observe that most of the explored configurations achieve a good tradeoff between DLA energy efficiency and GPU latency speedup. Furthermore, under the same baseline accuracy of \textit{Visformer}, we notice an energy gain up to $\sim$ \textbf{2.1x} compared to the GPU-only mapping with latency $\leqq 30ms$. Similarly, a latency speedup up to $\sim$ \textbf{1.7x} compared to the DLA-only mapping, with comparable energy efficiency. 
Secondly, we can notice an accuracy drop of $\sim$ 6\% when setting up hard constraints on the feature map reuse (See the \textit{50\% case}). Hence, defining the optimal inter-stages concatenation strategy that determines the feature maps reuse ratio is crucial to maintain the desired level of accuracy while minimizing inter-CUs dependencies.

%In fact, the MPSoC used in our evaluation allows sharing of data between CUs through a pinned memory limited by the availability of the DRAM. Nevertheless, including the constraint on the feature map reuse in the search process is an inevitable criterion for MPSoCs with limited shared memory. Thus, our experiments show that the more shared memory available for feature map reuse, the higher accuracy is achieved.

\subsection{Pareto Optimal Models Analysis}
%\blue{1) Two side-by-side barplots: One for the most energy-oriented model and one for the latency oriented model in each case. xticks: Acc, Ergy, Fmap reuse; categories: No constraint, 75\%, 50\%. (will probably have to explain what fmap reuse means in a sentence)}
%\blue{2)``We take a more in-depth look at the performance of the pareto-optimal models generated by the 3 search processes. Specifically, we focus on the most energy- and latency- oriented models from the frontier. In Figure \ref{barplot}, ..''}

\begin{figure}[]
\centering
  \includegraphics[width=0.5\textwidth]{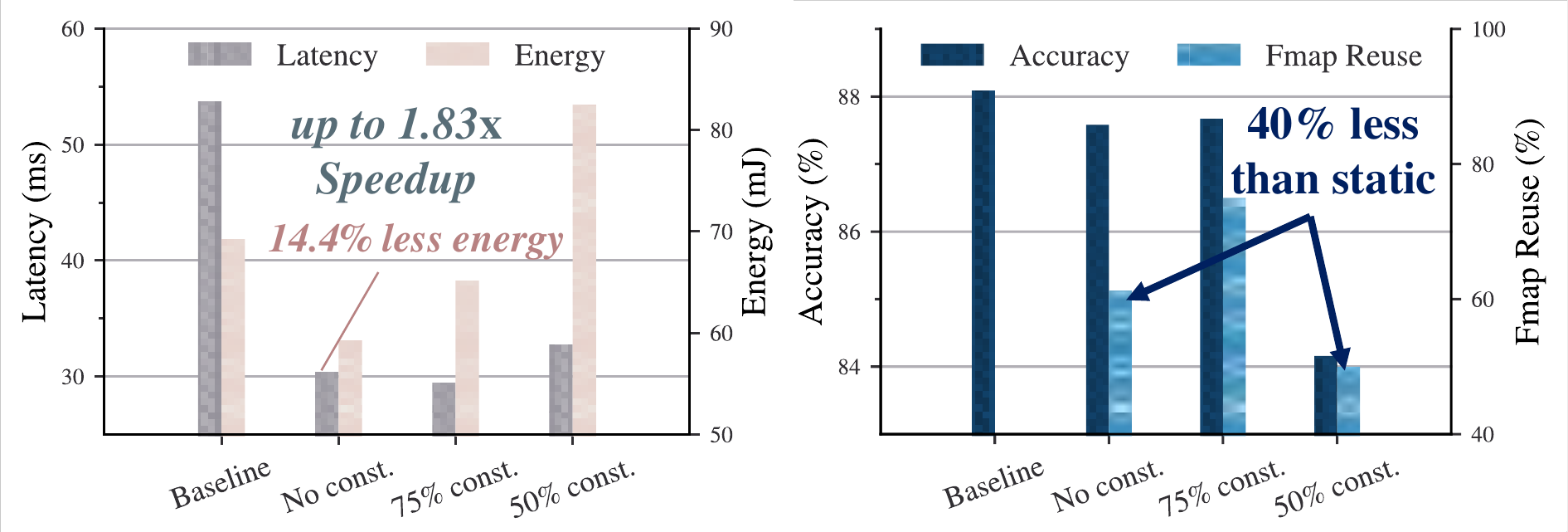}
  \vspace{-4.5ex}
  \caption{Comparison between the most energy-oriented models selected from the obtained Pareto sets by each search strategy and the baseline on DLA}
  \label{fig:pareto_models}
  \vspace{-3ex}
\end{figure}

In this section, we delve further into the performance breakdown of the Pareto optimal models obtained from the three search strategies. We select the most energy-oriented models and compare them with the baseline \textit{Visformer} mapped entirely on the DLA. Figure \ref{fig:pareto_models} and Table \ref{tab:perf_breakdown} detail the obtained results. By exploring neural network dynamicity and concurrency on heterogeneous CUs, our models achieve better latency-energy tradeoff, providing latency speedup of $\sim$ \textbf{1.83x} and up to $\sim$ \textbf{14.4\%} of energy gain as shown in the left sub-figure. In addition, the correlation between feature maps reuse and accuracy is highlighted in the right sub-figure. Reducing the feature maps reuse across stages decreases the inter-CUs data transmission at the cost of accuracy drops. 
%This is clearly shown, especially for the \textit{50\% constr.} with an accuracy drop of 6\%. 
However, some models can achieve comparable accuracy to the baseline while only reusing \textbf{60\%} of the necessary feature maps (See \textit{No constr. and 75\% constr.} cases) 

\subsection{Generalization to other architecture}
%\blue{1) Table II: but only: accuracy, Avg energy, avg latency, Fmap reuse }
%\blue{2) This can be a super small subsection tbh: ``Here, we demonstrate the applicability of our approach to a VGG-19 network as a representative of the standard convolutional neural networks class. We present our full set of results alongside those of the Visformer in Table \ref{}. ... ''  Highlihgt one two results/patterns of the VGG models in the architecture and that's it}
To further demonstrate our approach's applicability, we evaluate our optimization framework on a typical CNN architecture, \textit{VGG19}. Table \ref{tab:perf_breakdown} details the obtained results. Regarding the baseline performances, \textit{VGG19} depicts a high energy consumption on GPU and slow execution latency on DLA. This is explained by its many weights and large feature maps, which entail high memory footprints for both CUs. Moreover, the large number of weights may exhibit a high degree of redundancy. Our approach has exploited these two properties of \textit{VGG19} well, resulting in up to $\sim$ \textbf{4.62x} energy gain and $\sim$ \textbf{4.44x} latency speedup. Furthermore, according to our analysis, more than 80\% of samples were correctly classified in earlier stages with fewer channels, which results in considerable latency and energy gains. 

% And that concludes our experiments :)

\section{Conclusion}
%\blue{just summarize what the work presented and future work should aim to investigate extending our approach to other classes of architectures different heterogeneous MPSoC platforms}
% OD: No need (too detailed)
% extensibility to residual networks
% we find that the gpu for the initial stage is favorable for some reason
% Extension to hw/sw co-design
% \red{We should specify that we're not proposing an end-to-end system here and future work will be dedicated to integrate our approach into existing frameworks such as Jedi}
We have presented \textit{Map-and-Conquer}, an energy-efficient execution scheme for dynamic neural networks on heterogeneous MPSoCs by jointly optimizing the model partitioning along the width, hardware mapping, and \textit{DVFS}. \textit{Map-and-Conquer}'s awareness of the $\mathcal{NN}$ dynamicity and hardware computing units capabilities allows it to realize better performance trade-off over conventional single-platform mapping schemes. On CIFAR-100 and the AGX Xavier MPSoC,
\textit{Map-and-Conquer} achieved up to 2.1x energy gains over GPU-only mapping and up to 1.7x speedup over DLA-only mapping.

\bibliographystyle{IEEEtran}
\bibliography{references}

\end{document}